\documentclass[twocolumn,
               aps,
               prd,   
               preprintnumbers,
               numbers,
               sort&compress,
               nofootinbib,
               showpacs,
               colorlinks,
               linkcolor=blue,   
               citecolor=blue]{revtex4-1}

\usepackage{graphicx,amsmath,amssymb,bm}
\usepackage{psfrag}
\usepackage{feynmp}
\usepackage{hyperref}
\usepackage{graphicx}
\usepackage{subcaption}
\usepackage{placeins}
\usepackage{xcolor} 
\usepackage{gensymb}

\def\<{\langle}
\def\>{\rangle}

\def\+{\dagger}

\def\U1A{U(1)$_{\rm A}$}

\def\ra{\rangle}
\def\la{\langle}

\newcommand{\be}{\begin{eqnarray}}
\newcommand{\ee}{\end{eqnarray}}
\newcommand{\beq}{\begin{equation}}
\newcommand{\eeq}{\end{equation}}
\newcommand{\exclude}[1]{}
\newcommand{\jcap}{JCAP}

\newcommand{\physrep}{Physics Reports}

\newcommand{\kmps}{\,\mathrm{km}\,\mathrm{s}^{-1}}

\begin{document}

\title{Axion Quark Nuggets  and how  a Global Network    can discover them}
 
\author{Dmitry Budker}
       \email{budker@uni-mainz.de}
        \affiliation{Johannes Gutenberg-Universit{\"a}t Mainz (JGU) - Helmholtz-Institut, 55128 Mainz,   Germany\\
        Department of Physics,
University of California, Berkeley,  CA, 94720-7300, USA}
\author{Victor V. Flambaum}
       \email{v.flambaum@unsw.edu.au}
        \affiliation{School of Physics, University of New South Wales, Sydney 2052, Australia\\
        Johannes Gutenberg-Universit{\"a}tMainz (JGU) - Helmholtz-Institut, 55128 Mainz,   Germany}
\author{Xunyu Liang}
\email{xunyul@phas.ubc.ca}
 \author{Ariel Zhitnitsky}
\email{arz@phas.ubc.ca}
\affiliation{
 Department of Physics and Astronomy, University of British Columbia, Vancouver, Canada}
 
\label{firstpage}

\begin{abstract}

We advocate an idea that 
 the presence of the daily   and annual   modulations of the axion flux on the Earth’s surface may dramatically change the strategy of the axion searches. 
Our computations are based on the so-called Axion Quark Nugget (AQN) dark matter model
which was originally put forward to explain the similarity of the dark and visible cosmological matter densities $\Omega_{\rm dark}\sim \Omega_{\rm visible}$. In our framework, the population of galactic axions with mass  $ 10^{-6} {\rm eV}\lesssim m_a\lesssim 10^{-3}{\rm eV}$ and velocity $\la v_a\ra\sim 10^{-3} c$ will be  always accompanied by the axions with typical velocities $\la v_a\ra\sim 0.6 c$ emitted by AQNs.  We formulate the broadband detection strategy to search for such relativistic axions by studying the daily and annual modulations.  We describe several tests which could effectively  discriminate a true signal from noise.  These AQN-originated  axions can be observed  as correlated events  which could be recorded    by   synchronized    stations in the global network. The correlations can be effectively  studied if the detectors are positioned  at distances  shorter than a few hundred kilometres. 
  
\end{abstract}
\vspace{0.1in}


\maketitle

\section{Introduction} 
The Peccei-Quinn mechanism, accompanied by axions, remains the most compelling resolution of the strong   $\cal{CP}$ problem, see  original papers
\cite{1977PhRvD..16.1791P,1978PhRvL..40..223W,1978PhRvL..40..279W,KSVZ1,KSVZ2,DFSZ1,DFSZ2} and recent reviews \cite{vanBibber:2006rb, Asztalos:2006kz,Sikivie:2008,Raffelt:2006cw,Sikivie:2009fv,Rosenberg:2015kxa,Marsh:2015xka,Graham:2015ouw,Ringwald:2016yge,Battesti:2018bgc,Irastorza:2018dyq}. The  conventional idea for production of the dark matter (DM) axions  is either by the misalignment mechanism 
\exclude{\cite{1983PhLB..120..127P,1983PhLB..120..133A,1983PhLB..120..137D},} 
when the cosmological field $\theta(t)$ oscillates and emits cold axions before it settles at a minimum, or via the decay of topological objects, see   recent reviews \cite{vanBibber:2006rb, Asztalos:2006kz,Sikivie:2008,Raffelt:2006cw,Sikivie:2009fv,Rosenberg:2015kxa,Marsh:2015xka,Graham:2015ouw,Ringwald:2016yge,Battesti:2018bgc,Irastorza:2018dyq}.

\exclude{\cite{Chang:1998tb,2012PhRvD..85j5020H,2012PhRvD..86h9902H,Kawasaki:2014sqa,Fleury:2015aca,Gorghetto:2018myk,Klaer:2017ond}.   
}
 In addition to these well established mechanisms,  a fundamentally novel mechanism for axion production
was studied   in recent papers  \cite{Fischer:2018niu,Liang:2018ecs,Lawson:2019cvy,Liang:2019lya}.
This  mechanism is rooted in the so-called axion quark nugget (AQN) dark matter  model \cite{Zhitnitsky:2002qa}.
The AQN construction in many respects is 
similar to the original quark-nugget model suggested  by Witten \cite{Witten:1984rs}, see  \citep{Madsen:1998uh} for a review. This type of DM  is ``cosmologically dark'' not because of the weakness of the AQN interactions, but due to their small cross-section-to-mass ratio, which scales down many observable consequences of an otherwise strongly-interacting DM candidate. 

There are two additional elements in the  AQN model compared to the original proposal \cite{Witten:1984rs,Madsen:1998uh}. First, there is an additional stabilization factor for the nuggets provided by the axion domain walls which   are copiously produced  during the  QCD  transition which help to alleviate a number of  problems with the original \cite{Witten:1984rs,Madsen:1998uh} nugget model.\footnote{\label{first-order}In particular, a first-order phase transition is not a required feature for the nuggets' formation as the axion domain wall (with internal QCD substructure)  plays the role of the squeezer. Another problem with \cite{Witten:1984rs,Madsen:1998uh} is that nuggets likely evaporate on a Hubble time-scale. For the AQN model this is not applicable because the vacuum-ground-state energies inside (the color-superconducting phase) and outside (the hadronic phase) the nugget are drastically different. Therefore, these two systems can coexist only in the presence of an external pressure, provided by the axion domain wall. This should  be contrasted with the original model \cite{Witten:1984rs,Madsen:1998uh}, which must be stable at zero external pressure.}  Another feature of AQNs is that nuggets can be made of {\it matter} as well as {\it antimatter} during the QCD transition. The direct consequence of this feature is that  DM density, $\Omega_{\rm DM}$, and the baryonic matter density, $\Omega_{\rm visible}$, will automatically assume the  same order of magnitude  $\Omega_{\rm DM}\sim \Omega_{\rm visible}$ without any fine tuning.  This is because they have the same QCD origin  and are both proportional to the same fundamental dimensional parameter $\Lambda_{\rm QCD}$  which ensures that the relation $\Omega_{\rm DM}\sim \Omega_{\rm visible}$  
always holds irrespective of the parameters of the model such as the axion mass $m_a$ or  misalignment angle $\theta_0$. 

The existence of both AQN species   explains  the observed asymmetry between matter and antimatter as a result of separation of the baryon charge and generation of the disparity between matter and antimatter nuggets 
as a result of strong $\cal{CP}$ violation during the QCD epoch. 
Both AQNs with matter and antimatter   serve as dark matter in this framework.  In particular, if the number of anti-nuggets is larger than the number of nuggets by a factor of $3/2$ at the end of the formation, the ratio between visible and dark matter components assumes its observed value $\Omega_{\rm DM}\simeq 5~ \Omega_{\rm visible}$, while the total baryon charge of the Universe  (including the nuggets, anti-nuggets and the visible baryons)  remains zero at all times. 
This should be  contrasted with the conventional baryogenesis paradigm where extra baryons (1 part in $10^{10}$)  must be produced during the early stages of the evolution of the Universe to match the observations. 

We refer the reader to the original papers \cite{Liang:2016tqc,Ge:2017ttc,Ge:2017idw,Ge:2019voa} 
devoted to the specific questions  related to the nugget formation, generation of the baryon asymmetry, and how the nuggets survive the ``unfriendly'' environment of the early Universe.  Here we would like to make   several generic comments relevant for the present studies.   First,  
the AQN framework   resolves  two fundamental problems simultaneously: the nature of dark matter and the asymmetry between matter and antimatter.   Second, the AQNs are composite objects consisting of axion field  and quarks and gluons  in the color superconducting (CS) phase, squeezed by the axion  domain wall (DW).
This represents an absolutely stable  system  on cosmological time scales  as it assumes  the lowest-energy configuration for a given baryon charge. 
Third, while  the model was originally invented to explain the observed relation $\Omega_{\rm DM}\sim \Omega_{\rm visible}$  as mentioned above,  it may also explain a number of other (naively unrelated, but observed) phenomena, see below. 

The  AQNs may also offer a resolution to the so-called ``Primordial Lithium Puzzle" \cite{Flambaum:2018ohm}, the ``Solar Corona Mystery"  \cite{Zhitnitsky:2017rop,Raza:2018gpb}, and may also explain the recent EDGES observation \cite{Lawson:2018qkc}, which is in some tension with the standard cosmological model. Furthermore, it may resolve \cite{Zhitnitsky:2019tbh} the longstanding puzzle with the DAMA/LIBRA  observation \cite{Bernabei:2018yyw} of the annual modulation at $9.5\sigma$ confidence level, which is in direct conflict with other DM experiments if interpreted in terms of WIMP-nuclei interaction.    In the present studies we adopt the same set of physical  parameters of the model which were used in  explanation of the aforementioned phenomena. 

The key parameter which essentially determines all the intensities for the effects mentioned above is the average baryon charge $\la B \ra$ of the AQNs. There is a number of constraints on this parameter which are reviewed below. 
One should also mention that the AQNs masses related to their baryon charge by $M_N\simeq m_p |B|$, where we ignore small differences between the energy per baryon charge in  CS   and  hadronic confined  phases.  The resulting AQN are macroscopically large objects with a typical size of $R\simeq 10^{-5}{\rm cm}$ and roughly nuclear density resulting in masses   roughly 10\,g. For the present work we adopt a typical nuclear density of order $10^{40}\,{\rm cm^{-3}}$ such that a nugget with $|B|\simeq 10^{25}$ has a typical radius $R\simeq 10^{-5}{\rm cm}$.

The strongest direct detection limit 
is  set by the IceCube Observatory's non-detection of a non-relativistic magnetic monopole
\cite{Aartsen:2014awd}. While the magnetic monopoles and the AQNs interact with material of the detector
differently, in both cases the interaction leads to electromagnetic and hadronic cascades along the trajectory 
of AQN (or magnetic monopole) which must be observed by the  detector if such an event occurs.  A non-observation of any such cascades puts the following  limit on the flux of heavy nonrelativistic particles passing through the detector,   see Appendix A in \cite{Lawson:2019cvy}:
\be
\label{direct}
\la B \ra > 3\cdot 10^{24} ~~~[{\rm direct ~ (non)detection ~constraint]}.
\ee
 
Similar limits are also 
obtained from the Antarctic Impulsive Transient Antenna  (ANITA) \cite{Gorham:2012hy}.  In the same work the author also  derives the   constraint arising from a potential contribution of the AQN annihilation events to the Earth's energy budget 
requiring $|B| > 2.6\times 10^{24}$ \cite{Gorham:2012hy}, which is consistent with (\ref{direct}). 
 There is also a 
constraint on the flux of heavy dark matter with mass $M<55\,$g based on the non-detection of 
etching tracks in ancient mica \cite{Jacobs:2014yca}. It slightly touches the lower bound   (\ref{direct}), but does not strongly constrain the entire window  (\ref{window}).

The authors of \cite{Herrin:2005kb} use the Apollo data to constrain the abundance of  quark nuggets  in the region of 10\,kg to one ton. It has been argued that the contribution of such heavy nuggets  must be at least an order of magnitude less than would saturate the dark matter in the solar neighbourhood \cite{Herrin:2005kb}. Assuming that the AQNs do saturate the dark matter, the constraint  \cite{Herrin:2005kb} can be reinterpreted   that at least $90\%$ of the AQNs must have masses below 10\,kg. This constraint can be approximately expressed  in terms of the baryon charge:
   \be
\label{apollo}
\la B \ra \lesssim   10^{28} ~~~ [  {\rm   Apollo~ constraint ~} ]
\ee
Therefore, indirect observational constraints (\ref{direct}) and (\ref{apollo}) suggest that if the AQNs exist and saturate the dark matter density today, the dominant portion of them   must reside in the window: 
\be
\label{window}
3\cdot 10^{24}\lesssim\la B \ra \lesssim   10^{28}~ [{\rm constraints~ from~ observations}].  ~~~
\ee 
 
Completely different and independent observations  also suggest  that the galactic spectrum 
contains several excesses of diffuse emission the origin of which is not well established,  and  remains to be debated.
The best-known example is  the strong galactic 511~keV line. If the nuggets have a baryon 
number in the $\langle B\rangle \sim 10^{25}$ range they could offer a 
potential explanation for several of 
these diffuse components. It is a nontrivial consistency check that the required $\langle B\rangle$ to explain these excesses of the galactic diffuse emission  belongs to the same mass   range as stated above. 
 For further details see the original works \cite{Oaknin:2004mn, Zhitnitsky:2006tu,Forbes:2006ba, Lawson:2007kp,Forbes:2008uf,Forbes:2009wg} with explicit  computations 
 of the galactic radiation  excesses  for varies frequencies, including the observed excesses of the diffuse  x- and   $\gamma$- rays.  
In all these cases the intensity of the photon emission  is  expressed in terms of a single parameter $\langle B\rangle$ such that all relative intensities   are unambiguously fixed because  they are determined by the Standard Model (SM) physics.

Yet another AQN-related effect might be intimately linked to the so-called ``solar corona heating mystery".
The  renowned  (since 1939)  puzzle  is  that the corona has a temperature  
$T\simeq 10^6$K which is 100 times hotter than the surface temperature of the Sun, and 
conventional astrophysical sources fail to explain the extreme UV (EUV) and soft x ray radiation 
from the corona 2000 km  above the photosphere. Our comment here is that this puzzle  might  find its  
natural resolution with the same  baryon charge $\la B \ra$  from window (\ref{window}) which was   constrained from drastically different systems as reviewed above. 

We emphasize that the AQN model within window (\ref{window})  is consistent with all presently available cosmological, astrophysical, satellite and ground-based constraints. This model is  very rigid and predictive as there is no much flexibility nor freedom to modify any estimates  in different systems    as reviewed in this Introduction. In particular, the AQN-induced flux (\ref{flux0}) which plays a key role in the present studies cannot  change its numerical value for  more than factor of 2 , depending on the size distribution within the   window 
(\ref{window}). The same comment also applies to all other observables such as modulation parameters $\kappa_{\rm (a)}$ and  $\kappa_{\rm (d)}$ and  amplification factor $A (t) $ to be discussed in the present work.

\section{AQN-induced axion flux on Earth}\label{AQN-flux}
Relevant for the present studies   consequence of the construction is
  that  the axion portion of the energy   contributes to about  1/3 of the total AQN's mass in the form of the axion DW surrounding the nugget's core. 
This system  represents  a   time-independent configuration  which kinematically cannot convert its axion related energy (generated at earlier times during the QCD formation  epoch) to freely propagating time-dependent axions. However, any time-dependent perturbation, such as passage of the AQN  through the Earth's  interior,
inevitably results \cite{Liang:2018ecs}  in emission of real propagating relativistic axions with typical velocities $\la v_a\ra\sim 0.6c$ ($c$ is the speed of light), liberating the initially stored axion energy.     
The energy flux of the AQN-induced axions on the Earth surface was computed in \cite{Liang:2019lya} using full-scale Monte Carlo simulations accounting for all possible  AQN trajectories traversing the Earth:
\be
\label{flux0}
\la E_a\ra \Phi^{\rm AQN}_a\simeq 10^{14} \left[{\rm\frac{eV}{cm^2s}}\right], ~~~ \la E_a\ra\simeq 1.3\,m_a, 
\ee
where $E_a$ is the axion energy and $\Phi_a^{\rm{AQN}}$ is the AQN flux.
The rate (\ref{flux0})  includes   all types of AQN  trajectories inside  the Earth's interior: trajectories where AQNs hit the surface with incident angles close to $0^o$ (in which  case the AQN crosses the Earth core and exits at the opposite side of the Earth) as well as  trajectories where  AQNs just touch the surface  with incident angles close to $90^o$, in which case AQNs leave without much annihilation in the deep underground. 
The result of the  summation  over all these  trajectories can be expressed in terms of the average mass (energy) loss $\langle \Delta m_{\rm AQN} \rangle $ per AQN.   The same information can also be expressed in terms of the average   baryon-charge loss per nugget  $\langle \Delta B\rangle $   as  these two are directly related: $\langle \Delta m_{\rm AQN} \rangle \approx m_p\langle \Delta B\rangle $, see  \cite{Liang:2019lya} for details. Let us repeat again: the expression  (\ref{flux0})  represents the average  flux accounting for different trajectories and AQN size distributions averaged over times much greater than a year.

For the purposes of the present work, it is important  to consider the time dependent modulation and amplifications effects which can be represented as follows: 
\be
\label{flux}
\la E_a\ra \Phi^{\rm AQN}_a(t)\simeq 10^{14}A(t) \left[{\rm\frac{eV}{cm^2s}}\right], ~~~ \la E_a\ra\simeq 1.3\,m_a, ~~
\ee
where $A(t)$ is the modulation/amplification time dependent  
factor. The factor $A$ for the daily and annual modulations is discussed in Sec.\,\ref{strategy} below and is given by Eqs.\,(\ref{eq:annual}) and 
(\ref{eq:daily}), correspondingly. In both cases, the factor $A$ does not deviate from the average value by more than $10\%$. However, sometimes the factor $A$
can be numerically large  for rare  bursts-like events,  
the so-called ``local flashes'' in the terminology  of Ref.\,\cite{Liang:2019lya}.   These  short  
bursts  (with a duration time of the order of a second for $A\simeq 10^2$ \cite{Liang:2019lya}) resulting from the interaction of the AQN hitting  the Earth in  a close vicinity of a detector. Another feature of the AQN induced axions distinguishing them from conventional galactic axions is that the typical velocities of the  AQN induced axions  are relativistic
 with $\la v_a\ra \sim 0.6 c$, in contrast to the  galactic axions with $\la v_a\ra \sim 10^{-3} c$.

 \begin{table} 
\captionsetup{justification=raggedright}
 	\caption{Estimations of Local flashes for different $A$ as defined by (\ref{flux}).  The corresponding event rate  and  the time duration $\tau$ depend on factor  $A$, which itself is determined by the shortest distance from the nugget's trajectory to the detector. The table is  adopted from \cite{Liang:2019lya}: } 
	\centering 
	\begin{tabular}{ccc}
		\hline \hline
		$A$ &  $\tau$ (time span) & event rate \\ 
		\hline 
		1 & 10 s & 0.3 $\rm min^{-1}$ \\ 
		$10$ & 3 s & 0.5 $\rm hr^{-1}$ \\ 
		$10^2$ & 1 s & 0.4 $\rm day^{-1}$ \\ 
		$10^3$ & 0.3 s & 5 $\rm yr^{-1}$ \\  
		$10^4$ & 0.1 s & 0.2 $\rm yr^{-1}$ \\  
		\hline 
	\end{tabular}
	\label{tab:local flashes}
\end{table}

It is instructive to compare the AQN-induced flux (\ref{flux}) with the flux computed from assumption that the galactic axions saturate the DM density $\rho_{DM}\sim 0.3\,{\rm GeV\cdot cm^{-3}}$ today. This assumption cannot be satisfied in the entire window of $ 10^{-6} {\rm eV} \lesssim m_a\lesssim 10^{-3} {\rm eV}$ as the conventional contribution is highly sensitive to $m_a$
as $\rho_{\rm DM} \sim m_a^{-7/6}$ and may saturate the DM density at $m_a\lesssim 10^{-5} {\rm eV}$, depending on additional assumptions on production mechanism. It should be contrasted with the AQN framework where $\Omega_{\rm DM}\sim \Omega_{\rm visible}$  always holds irrespective of the parameters of the model such as the axion mass $m_a$ or  misalignment angle $\theta_0$.  This, in particular, implies that for $m_a\gtrsim 10^{-4} {\rm eV}$ the conventional galactic axions contribute very little to $\Omega_{\rm DM}$ while the AQNs are the  dominant contributor to the DM density. Nevertheless, in what follows we need a point of normalization with conventional picture and conventional estimates. With this purpose in mind, here and in what follows we compare the AQN induced flux (\ref{flux}) with $A=1$ with  conventional galactic axion flux computed with the assumption formulated above. In this case  the numerical value for the flux (\ref{flux}) is approximately two orders of magnitude below the value computed for the  conventional galactic axions. 

The cavity type experiments such as ADMX are to date the only ones to probe the parameter space of the conventional QCD axions with $\la v_a\ra \sim 10^{-3} c$, while we are interested in detection of the relativistic axions with  $\la v_a\ra \sim 0.6 c$. This requires a different type of instruments and drastically different search strategies. We argue below that the daily and annual modulations (\ref{eq:annual}) and 
(\ref{eq:daily}) as well as the short bursts-like amplifications with $A\simeq 10^2$ might be the key elements in formulating a novel  detection strategy to observe these effects, which is precisely the topic of the present work. 
 
Let us reiterate  that the  goal of the present work is not to design a specific  instrument which would be  capable of detecting 
the axions being emitted by AQNs and would be the sensing element of the synchronized  stations assembled in a global network. For example, the presently operating  Global Network of Optical Magnetometers for Exotic physics searches (GNOME)  \cite{2013arXiv1303.5524P,Afach:2018eze} is sensitive to frequencies of up the kHz range, while the preferred value for the axion mass for the AQN dark matter is $m_a\simeq 10^{-4}{\rm eV}$ corresponding to 24 GHz.  
  
The present work is devoted to a completely different question. We wish to develop a strategy which would provide a future framework to study the axions emitted by AQNs. While   there are no presently  available instruments  operating in the  interesting window: $ 10^{-6} {\rm eV} \lesssim m_a\lesssim 10^{-3} {\rm eV}$  we do not see any fundamental  obstacles  which would prevent designing and building the required instruments  in future.   
 In what follows we {\it assume} that the axion search detectors sensitive to  24 GHz can be designed and built, for example 
 using single-photon detectors  for the GHz range  \cite{Kuzmin2013,Lamoreaux:2013koa}. 

There are several key ingredients  in our  proposal. First of all, as already mentioned, the secondary axions emitted by AQNs are relativistic with $\la v_a\ra \sim 0.6 c$, in contrast to conventional galactic axions with $\la v_a\ra \sim 10^{-3} c$. This has an important implication for the proposed search because the axion is broadband with $\Delta \nu/\nu\sim 1$, in contrast with conventional narrow-line galactic axions with $\Delta \nu/\nu\lesssim 10^{-6}$ searched for with the cavity-type detectors. Second, 
we {\it assume} that a GNOME-like network sensitive to the required frequencies and   spectral features can be built in the future. The strategy for detecting broadband axions is formulated  in  Sec.\,\ref{strategy}. 
  
\exclude{
The axion production  mechanism in the AQN framework has been developed in  \cite{Fischer:2018niu,Liang:2018ecs,Lawson:2019cvy}. Therefore, to avoid any repetition of the background  material  we refer to that papers  regarding the computations of the spectrum and  intensity. We refer to \cite{Liang:2019lya} devoted to  specific    questions related to the time modulations and amplifications which always accompany the passage of the AQNs through the Earth. This part of the analysis   represents  the key element for the present proposal   as the daily and annual modulations may help drastically reduce the noise, while  the observation of the  ``local flashes" using a global network   with large amplification factor $A\sim (10^2-10^4)$ may unambiguously pinpoint the relevant frequency bin in a broadband detection search where the modulation is observed, see Section \ref{strategy}.
}

\section{Basic Idea, notations and definitions}\label{idea}
The starting point of our analysis is the Hamiltonian describing the coupling of the spin operator (for electrons or nucleons) with the gradient of the axion field. The same coupling was discussed for the CASPEr experiment \cite{Graham:2013gfa,Wu2019,Garcon2019}  in the case of nucleons and for QUAX \cite{Barbieri:2016vwg}  in the case of electrons. This coupling
is analogous to the Zeeman effect (the basis of magnetometry \cite{Budker:2006gya}) with the gradient of the pseudoscalar $\boldsymbol{\nabla} a(\mathbf{r}, t)$ being a pseudovector analogous to magnetic field:
\be
\label{H}
H_{\rm spin}\simeq g_{\rm a} \boldsymbol{\sigma}\cdot \boldsymbol{\nabla} a(\mathbf{r}, t),~~~ g_{\rm a }\propto f_a^{-1}.
\ee
Here, the coupling constant $g_a$ assumes the value $g_{\rm a }\equiv g_{\rm aee}$ for electrons or $g_{\rm a }\equiv g_{\rm aNN}$ for nucleons in notations of Ref.\,\cite{Graham:2013gfa} and $f_a$ is the so-called axion decay constant. 
The coupling (\ref{H}) describes the interaction of the spins of a material with an oscillating pseudo-magnetic field $\mathbf{B}_a\propto\boldsymbol{\nabla} a(\mathbf{r}, t)$ generated by the gradient of the propagating axion $a(\mathbf{r}, t)=a_0\exp(-iE_a t+i\mathbf{p}_a\cdot\mathbf{r})$, where the normalization constant $a_0$  can be expressed in terms of the AQN-induced flux (\ref{flux}) computed on the Earth's surface, see below.
The maximum magnitude of the perturbation due to the coupling (\ref{H}) can be estimated as 
\be
\label{Delta}
 {\Delta E} \simeq  g_{\rm a} m_a a_0(\boldsymbol{\sigma}\cdot \mathbf{v}_a)\sim 10^{-8}\sqrt{A}~ {\rm s}^{-1}\left(\frac{g_{\rm a}}{10^{-9}~ {\rm GeV}^{-1}}\right), ~~~
\ee
where we estimated normalization  factor $a_0$ using AQN-induced flux (\ref{flux}). In conventional energy  units, ${\Delta E}\simeq 6\cdot 10^{-24} \sqrt{A}~ {\rm eV}$.
The strength of the interaction (\ref{H}) is normally expressed in terms of the pseudo-magnetic field $B_a$ which for
nucleon and electron systems assumes the following values:
\be
\label{B}
B_a^{N}&\equiv& \frac{\Delta E}{\mu_N}\simeq 2\cdot 10^{-16} \sqrt{A}\left(\frac{g_{\rm aNN}}{10^{-9}~ {\rm GeV}^{-1}}\right)~~ {\rm T}, \nonumber\\  B_a^{e}&\equiv& \frac{\Delta E}{\mu_e}\simeq   10^{-19}  \sqrt{A}\left(\frac{g_{\rm aee}}{10^{-9}~ {\rm GeV}^{-1}}\right)~~{\rm T}.
\ee
It is instructive to compare our estimate (\ref{Delta})  for the AQN-induced axions with similar estimate for the conventional galactic axions saturating the galactic DM density. As one can see from (\ref{Delta})  the numerical value for $\Delta E$ [and correspondingly for $B_a$ given by (\ref{B})] is approximately three times larger for the AQN-induced axions (in comparison with corresponding estimate of Ref.\,\cite{Graham:2013gfa} for galactic axions) even without amplification  due to  two effects working in opposite direction. The AQN-induced axion flux is 
two orders of magnitude smaller than the  galactic axion flux.  As typical axion galactic velocities are $10^{-3}$c, while the AQN-induced axions are relativistic with  $\la v_a\ra \simeq 0.6 c$, the corresponding   AQN-induced axion density is five orders of magnitude smaller than the galactic axion density.
As $\Delta E$ depends on the axion density  as $\sqrt{n_a}$ this gives a suppression factor $\sqrt{10^{-5}}\sim 3\cdot 10^{-3}$ in comparison with estimates for the galactic axions.  However, the velocities of the AQN-induced axions are relativistic with $v_a\sim c$ which provides the enhancement factor $10^{3}$ as velocity  linearly enters (\ref{Delta}), which explains why $\Delta E$ given by (\ref{Delta}) is three times of the corresponding estimate  \cite{Graham:2013gfa}. The amplification factor $A$   makes  this enhancement even stronger. 
 
 A few comments are in order.  First of all, the observable (\ref{Delta}) as well as  the pseudo-magnetic field (\ref{B}) 
 depend on the amplitude of the axion field $a_0$, not on its intensity $n_a\sim |a_0|^2$.  This   implies that the signal 
 will show  the oscillating features with the frequency determined by $m_a$.
 
 Second, the axion field $a(\mathbf{r}, t)$ can be treated as a classical field because the number of the AQN-induced axions (\ref{flux}) accommodated by a single de-Broglie volume is large in spite of the fact that the de-Broglie wavelength $\lambda$ for relativistic AQN-induced axions  is much shorter than for galactic axions:
 \be
\label{density}
n_a^{\rm AQN}\lambda^3\sim \frac{\Phi_a^{\rm AQN}}{v_a}\cdot \left(\frac{\hbar}{m_a v_a}\right)^3\sim 10^6 \left(\frac{10^{-4} {\rm eV}}{m_a}\right)^4\gg 1. \nonumber
 \ee
We emphasize that the wavelength $\lambda$ of the emitted axions is   short, measured in centimetres, while the distance $\Delta R$ (relevant for detecting a correlation) between  the network  stations is measured in hundred kilometres. To reiterate: we are suggesting to study the {\it correlation} between the transient signals which could be detected by different network stations. It should be contrasted with a proposal to study the   {\it coherent} signal when the amplitude $a_{\rm ALP}$ of  axion  light particles (ALPs) with very small mass $m_{ALP}\simeq [10^{-12}-10^{-14}] ~{\rm eV}$  has a coherence length  scale  $\lambda_{ALP}\equiv m_{ALP}^{-1}\simeq [10^2-10^4]$ kilometres. The study of these ALPs is  not a topic  of the present work  as the axions being discussed here are exclusively conventional QCD  axions with a mass range of ($10^{-6} {\rm eV} \lesssim m_a\lesssim 10^{-3} {\rm eV}$) with short wavelength  measured in centimetres.
 
The final and most important for this work comment  is as follows. If there is a global network (GN) of axion-search detectors,  there will be a correlated signal which can be detected with several synchronized  
 GN  stations due to the ``local flash" from  {\it one and the same}  AQN traversing in close vicinity of these stations.  
 The corresponding correlations discussed in Section \ref{delays} play a  key role  in the  formulation of our novel detection strategy  because these correlations can unambiguously   remove ``fake" signals from the AQN-related events. 
 
 The presence of the daily and annual modulations  \cite{Liang:2019lya} of the axion flux on the Earth's surface along with the large average velocities   $\la v_a\ra \simeq 0.6 c$ of the emitted axions by AQNs  dramatically changes   entire  strategy of axion searches, the topic 
discussed in the next Sec.\,\ref{strategy}. After we explain the broadband detection strategy,  we turn to  Sec.\,\ref{delays}  
 where we present the arguments  suggesting that the most efficient configuration for our purposes is the presence of a subset 
 of several  GN stations which are positioned in close vicinity of each other with $\Delta R\sim 10^2$ km or less. 
 \exclude{
 This is because this subset of the GN  stations will be the subject of the amplification factor $A \sim (10^2-10^4)$
 produced by a {\it single} AQN traversing in close vicinity of these stations. Amplification factor $A\sim 10^2$ is expected to occur once a day in every  point on the earth surface, while stronger amplifications are relatively rare events   \cite{Liang:2019lya}.
 In this case the correlated signal can be observed by this subset of the GN  stations  with time delays $\Delta t\sim \Delta R/v_{\rm AQN}\sim 1{\rm\,s}$ where  $v_{\rm AQN}\sim 10^{-3}c$ is a typical velocity for DM particles. The time  duration of the AQN-induced  burst will assume a similar magnitude $\Delta \tau\sim 1\,$s, see next section with relevant estimates.} 
 
 \section{Detection of broadband axions}\label{strategy}
 
As the axions emitted by AQN have relativistic velocities with a large dispersion \cite{Liang:2018ecs}, the corresponding signal is expected to be spectrally broad. It should be contrasted with the conventional galactic axions searched for, for instance, in experiments based on tuning of the resonant frequency of a cavity to match the microwave photons produced by the axions in the presence  of a strong magnetic field. In the latter, one assumes that the galactic-DM axion velocities and their dispersion  are small $\delta v/c \sim \la v_a \ra /c\sim 10^{-3}$. The cavity type experiments  such as ADMX, ADMX-HF \cite{Stern:2016bbw},    HAYSTAC \cite{Zhong:2018rsr}, and the experiments at CAPP reviewed in \cite{Battesti2018} are  to date the  only experiments to probe the particularly interesting region of parameter space corresponding to standard QCD axion models with $ 10^{-6} {\rm eV} \lesssim m_a\lesssim 10^{-3} {\rm eV}$. The galactic axions generate a  narrow microwave resonance with $\Delta \nu/\nu\sim\left( \delta v/c\right) ^2 \sim 10^{-6}$ such that the cavity-type experiments are designed to search for such a narrow line. 

Since the photons produced by the axions from AQNs are broadband, with $\Delta \nu/\nu\sim 1$, one needs to use a correspondingly broadband detector and the conventional cavity detectors which are designed to search for narrow lines  should be replaced with broadband instruments
  such   ABRACADABRA \cite{Kahn:2016aff}, LC Circuit \cite{Sikivie:2013laa}, see also \cite{alex2018sensitivity},  
which detect axion-induced magnetic fields and can be operated in a broadband mode.   
The search strategy has to be correspondingly adapted for AQN induced axions.

An important specific feature of the spectrum of the AQN induced axions that can be used for discriminating against spurious signals is that it has a peak around $v_a\simeq 0.6c$ with a sharp cutoff at higher velocities around $v_a\geq 0.8c$ and a strong suppression at low velocities $v_a\lesssim 0.2c$, see Fig. 1a in \cite{Liang:2018ecs}. These features correspond to the axion frequency band as follows:
$m_a\leq \omega_a\leq 1.8~m_a$.

 
While there are presently no broadband experiments operating in the  interesting window: $10^{-6}\,{\rm eV} \lesssim m_a\lesssim 10^{-3}\,{\rm eV}$  we do not see any fundamental  obstacles  which would prevent one from designing and building a required instrument in the future.   
In what follows we {\it assume} that detectors sensitive to broadband axions can be designed and built.  
  
With this assumption in mind, a strategy to probe the QCD axion can be formulated as follows. It has been known since \cite{Freese:1987wu}   that the DM flux shows  annual modulation due to the differences in  relative orientations  of the DM wind and the direction of the Earth motion around the Sun. The corresponding effect for AQN induced  axions was
 computed in  \cite{Liang:2019lya}.   The daily modulation which is a feature for the AQN model 
was also computed in the same paper\footnote{Daily modulations are also present in galactic-axion ``wind'' experiments such as those of Refs.\,\cite{Wu2019,Garcon2019,Smorra2019}.}.  The broadband strategy is to separate a large frequency  band into a number of smaller frequency bins  with the width $\Delta \nu \sim \nu$ according to the axion dispersion relation as discussed above. 
 
The time dependent signal in each frequency bin $\Delta \nu_i$ has to be fitted according to the expected modulation pattern, daily, or annual. For example, the annual modulation should be fitted according to the following formula 
  \begin{equation}
\label{eq:annual}
 A_{\rm (a)}(t)\equiv[1+\kappa_{\rm (a)} \cos\Omega_a (t-t_0)],
\end{equation}
 where $\Omega_a=2\pi\,\rm yr^{-1}\approx 2\pi\cdot\,32\,$nHz is the angular frequency of the annual modulation
 and   label $``a"$ in $\Omega_a $ stands for annual.   The $\Omega_a t_0$ is the phase shift corresponding to the maximum on June 1 and minimum on December 1 for the standard galactic DM distribution, see \cite{Freese:1987wu,Freese:2012xd}. 
 
 The same procedure should be repeated for all frequency bins ``$i$".
 Let us assume that the modulation has been recorded in a specific bin $\bar{i}$. 
 The modulation coefficient $\kappa^{\bar{i}}_{\rm (a)}$ for a specific  $\bar{i}$ could be as large as $10\%$. The parameters $\Omega_a$, $\kappa^{\bar{i}}_{\rm (a)}$ and $t_0$ are to be extracted from the fitting analysis and compared with theoretical predictions. 
 
 A test that it is not a spurious signal is a relatively simple procedure: one should check that no modulations appear in all other bins (except to possible neighbours to  $\bar{i}$ bin). A more powerful  test to exclude a spurious signals is described in next section \ref{delays}. 
 One should comment here that  precisely this strategy has been used by the DAMA/LIBRA collaboration which has been observing  the annual modulation for 20 years\footnote{DAMA/LIBRA collaboration claims \cite{Bernabei:2018yyw} the observation for an annual modulation  in the $(1-6)~ {\rm keV}$ energy range  at $9.5 \sigma$ C.L. 
 The C.L. is even higher ($12.9 \sigma$) for $(2-6)~ {\rm keV}$ energy range  when DAMA/NaI and DL-phase1
 are combined with DL-phase2 results.  The measured period   ($0.999\pm 0.001$) year and phase corresponding to $t_0=145\pm 5$ days corresponding to the maximum of the signal around June 1.}. It is considered as a strong    evidence of the 
 dark-matter origin of the modulation
  for recoil energy   in bins $E_{\rm recoil}\simeq (1-6)~ {\rm keV}$, while the modulation vanishes outside this range, see the latest results in   \cite{Bernabei:2018yyw} and an explanation within AQN framework in \cite{Zhitnitsky:2019tbh}. 

 A similar procedure can be applied for the daily modulations and  can be described as follows \cite{Liang:2019lya}, 
 \begin{equation}
\label{eq:daily}
A_{\rm (d)}(t)\equiv[1+\kappa_{\rm (d)} \cos(\Omega_d t-\phi_0)],
\end{equation}
 where   $\Omega_d=2\pi\,\rm day^{-1}\approx 2\pi\cdot 11.6\,\mu$Hz is the angular frequency of the daily modulation, while $\phi_0$ is the phase shift similar to $\Omega_at_0$ in (\ref{eq:annual}). It   can be assumed to be constant on the scale of days. However, it actually slowly changes  with time due to the  variation of the direction of  DM wind   with respect  to the Earth. 
 
 In summary, the axions characterized by broad distribution with $m_a\leq \omega_a\leq 1.8~m_a$ as discussed above  
    will produce nonzero modulation coefficients $\kappa_{\rm (a)}$ and $\kappa_{\rm (d)}$  in one frequency  bin  $\bar{i}$ (or perhaps two neighbouring bins). It is a nontrivial consistency test that the modulation occurs in one and the same frequency bin $\bar{i}$ for two drastically different analyses: the fittings for (\ref{eq:annual}) and (\ref{eq:daily}), correspondingly. A further consistency check is see whether the modulation is observed in other frequency bins. 
    A   more sophisticated, but at the same time,  more powerful test is described below. 
    The next section should be considered as a powerful tool which discriminates the  true signal contributing to  (\ref{eq:annual}) and (\ref{eq:daily}) from a spurious noise background.

\section{Time delays and durations}\label{delays} 
In this section   we describe a test which would unambiguously suggest if the observed modulations is due to the noise and/or  systematic errors, or it represents a   truly  DM signal. The test is based on analysis of ``local flashes" which are burst like events.  

The mechanism of a local flash is the following: the flux of AQN-induced axions gains a large amplification factor $A$ in an instant when a moving AQN is sufficiently close to the detector, namely \cite{Liang:2019lya}
\begin{equation}
\label{eq:A(d)}
A(d)
\simeq\left(\frac{0.2R_\oplus}{d}\right)^2
=\left(\frac{1.27\times10^3{\rm\,km}}{d}\right)^2\ ,
\end{equation}
where $d$ is the shortest distance from the AQN to the detector, while $R_{\oplus}$ is the Earth's radius. The time duration of the local flash is by definition:
\begin{equation}
\label{eq:Delta tau}
\Delta\tau\equiv\frac{d}{v_{\rm AQN}}
\simeq4.25\,A^{-1/2}{\rm\,s}
\left(\frac{300\kmps}{v_{\rm AQN}}\right).
\end{equation}
Therefore, for amplification $A\gtrsim10^2$ the required distance from the detector to AQN is $d\lesssim10^2$ km. Consequently, for two nearby  GN stations located   $10^2$ km (or less) apart  there is a large chance to detect a correlated signal amplified by $A\sim10^2$ from \textit{one and the same} AQN.

\begin{figure}[h]
	\captionsetup{justification=raggedright}
	\includegraphics[width=0.8\linewidth]{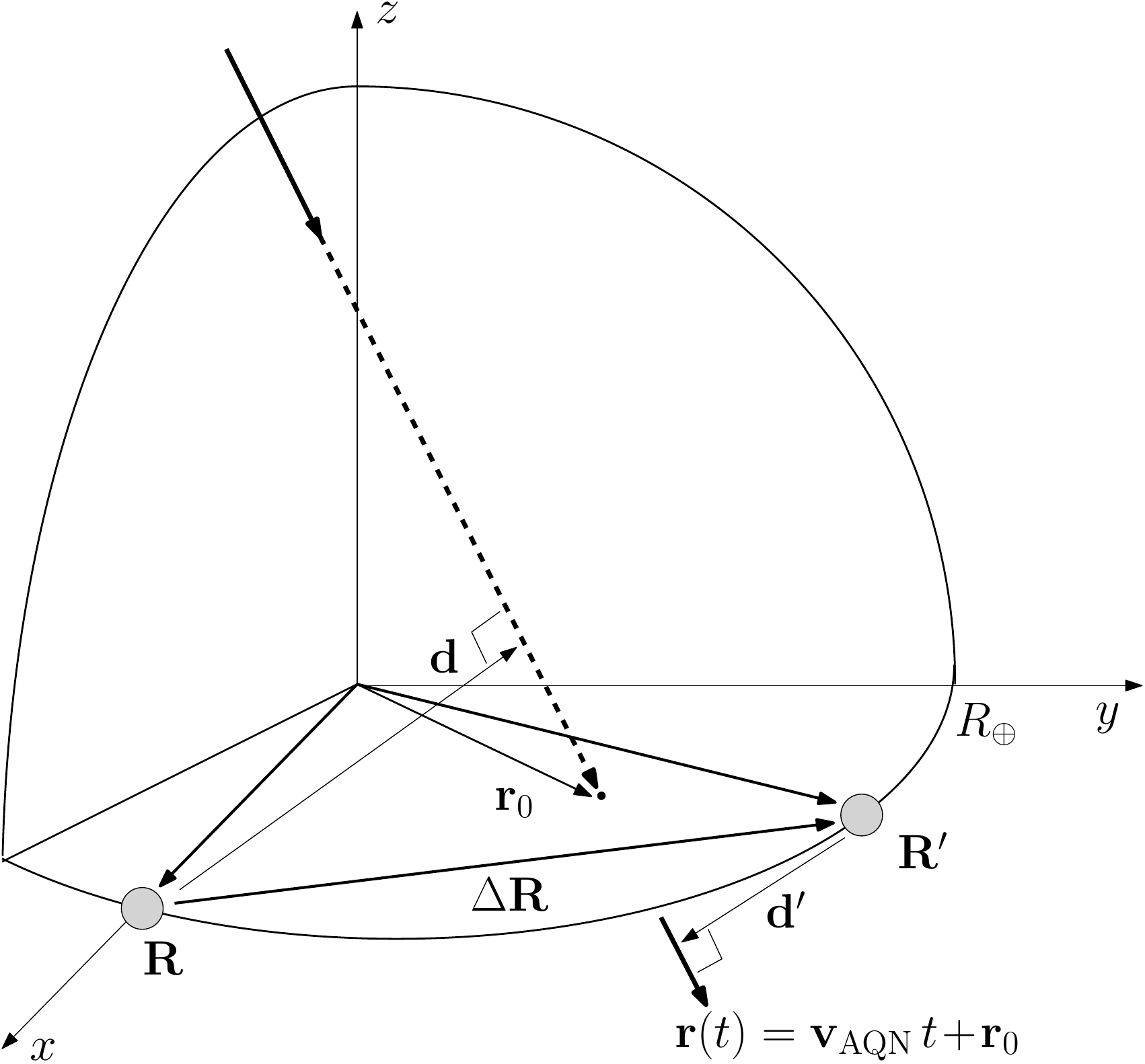}
	\caption{The two stations located at $\mathbf{R}$ and $\mathbf{R}'$ on the surface of Earth respectively. Each station has a distance $\mathbf{d}$ (and $\mathbf{d}'$) from the AQN trajectory $\mathbf{r}(t)=\mathbf{v}_{\rm AQN}\,t+\mathbf{r}_0$.}
	\label{fig:coordinate}
\end{figure}

To assess the time delay of a correlated signal, consider two stations located at $\mathbf{R}$ and $\mathbf{R}'$ on the surface of the Earth respectively, see Fig. \ref{fig:coordinate}. Now the first station detects a local flash when an AQN passes    nearby. The trajectory of the AQN is linear \cite{Lawson:2019cvy,Liang:2019lya} and can be described as:
\begin{equation}
\label{eq:bold r}
\mathbf{r}(t)
=\mathbf{v}_{\rm AQN}\,t+\mathbf{r}_0\ ,
\end{equation}
where $\mathbf{v}_{\rm AQN}$ can be approximated as a constant within the short time of correlated local flash $\sim1$ s, $\mathbf{r}_0$ is the intercept at the plane spanned by $\mathbf{R}$ and $\mathbf{R}'$. The distances from the stations to the AQN trajectory are denoted as $\mathbf{d}$ and  $\mathbf{d}'$ respectively. 

By imposing the orthogonal condition of $\mathbf{d}$ (and $\mathbf{d}'$) to $\mathbf{v}_{\rm AQN}$, we solve for the moment $t_*$ (and $t_*'$) when a peak signal of the local flash is detected in each station:
\begin{equation}
\label{eq:bold d dot bold v_AQN}
\begin{aligned}
&0=\mathbf{d}\cdot\mathbf{v}_{\rm AQN}
=[\mathbf{r}(t_*)-\mathbf{R}]\cdot\mathbf{v}_{\rm AQN},\\
&0=\mathbf{d}'\cdot\mathbf{v}_{\rm AQN}=[\mathbf{r}(t_*')-\mathbf{R}']\cdot\mathbf{v}_{\rm AQN}\, .
\end{aligned}
\end{equation}
The solutions give the time delay between two stations
\begin{equation}
\label{eq:Delta t}
\Delta t
\equiv |t_*'-t_*|
=\frac{\Delta R}{v_{\rm AQN}}\delta,\quad
\delta\equiv |\Delta\mathbf{\hat{R}}\cdot\mathbf{\hat{v}}|\ ,
\end{equation}
where $\Delta\mathbf{R}=\mathbf{R}'-\mathbf{R}$ is the separation distance between the two stations, as presented in Fig. \ref{fig:coordinate}. In practice, $\delta\in(-1,1)$ will be a free tuning parameter because the incident direction $\mathbf{\hat{v}}$ of the AQN trajectory is unknown. Assuming $\Delta R\sim10^2$ km and $v_{\rm AQN}\sim300\kmps$, we expect $\Delta t$ is no greater than $\sim 1$ s. For smaller $\Delta R$ the time delay $\Delta t$ decreases correspondingly. In particular, two detectors localized in the same building must  show  the synchronized pulses with zero time delay. 

One important relation in what follows can be derived from Eqs. \eqref{eq:bold d dot bold v_AQN} and \eqref{eq:Delta t}:
\begin{subequations}
\begin{equation}
\label{eq:bold d prime}
\mathbf{d}'
=\mathbf{d}+\mathbf{v}_{\rm AQN}\Delta t-\Delta\mathbf{R}\ ,
\end{equation}
\begin{equation}
\label{eq:abs d}
d'
\equiv|\mathbf{d}'|
\leq d+\Delta R (1+\delta)\ .
\end{equation}
\end{subequations}
Here Eq. \eqref{eq:bold d prime} can be also understood directly from the vector configuration in Fig. \ref{fig:coordinate}, and Eq. \eqref{eq:abs d} is based on the inequality $|\mathbf{a}+\mathbf{b}|\leq|\mathbf{a}|+|\mathbf{b}|$.  

To ensure a correlated signal distinguishable from background noise, amplifications received in both stations need to be sufficiently large. Assuming a local flash is detected in the first station with amplification $A(d)$, the constraint to the second station is clearly $d'\lesssim d$ or, according to Eqs. \eqref{eq:A(d)} and \eqref{eq:abs d}:
\begin{equation}
\label{eq:Delta R lesssim}
\Delta R
\lesssim\frac{d}{1+\delta}
\simeq 85{\rm\,km}\left(\frac{1.5}{1+\delta}\right)
\left(\frac{10^2}{A}\right)^{1/2}\ ,
\end{equation}
where $A\equiv A(d)$ for brevity of notation, and $\delta\simeq0.5$ is estimated by assuming a uniform distribution of AQN flux. Hence, to observe a correlated signal from two nearby stations with amplification $A\gtrsim10^2$, the   separation distance should be  85 km or less. 

Lastly, we estimate the event rate of a correlated signal for a given amplification $A$. The event rate for a single station has been estimated in Ref. \cite{Liang:2019lya}. The correlated event rate (CER) is the single event rate multiplied by an additional suppression factor (as presented in square bracket below):
\begin{equation}
\label{eq:CER}
\begin{aligned}
{\rm CER}
&\sim 0.29 A^{-3/2}{\rm\,min}^{-1}
\left[\frac{\frac{1}{2}\pi d^2}{2\pi \Delta R^2}\right]  \\
&\gtrsim 0.23{\rm\,day}^{-1}\left(\frac{1+\delta}{1.5}\right)^2
\left(\frac{10^2}{A}\right)^{3/2}\ .
\end{aligned}
\end{equation}
\exclude{
In the first step, the suppression factor is estimated by the ratio of cross section as shown in Fig. \ref{fig:cross section}. The following step estimates the lower limit of the CER by inequality \eqref{eq:Delta R lesssim}.
}
 Comparing to the single event rate calculated in Ref. \cite{Liang:2019lya}, the CER is suppressed by roughly one half for two nearby stations subject to constraint \eqref{eq:Delta R lesssim}. 
 
 We conclude this section with the following remark. 
 The AQN model unambiguously predicts the intensity of the flux (\ref{flux}) with well-defined amplification  parameters $A$ listed in Table \ref{tab:local flashes}. As mentioned above, there is no
specific instrument at this time that is sensitive to the  relevant frequency band 
and which could  effectively use  the broadband detection strategy as described in this paper.  Therefore, we cannot 
estimate the relevant sensitivity of an instrument at this point. However, such  estimations can be performed in the future as the basic physics parameters  such as the flux   (\ref{flux}) and  the modulation   parameters (\ref{eq:annual}) and  (\ref{eq:daily}) are unambiguously fixed in this framework, and there is no room nor flexibility to modify them.  
 
 \section{Conclusion} The presence of the daily (\ref{eq:daily}) and annual (\ref{eq:annual}) modulations of the axion flux on the Earth’s surface along with the large average velocities of the  axions emitted by AQNs dictates the search strategy for such axions. We suggest broadband detection
 to attack this problem as described in Section \ref{strategy}. We also suggest several tests to discriminate the DM signal from spurious signal.
  A sophisticated and  powerful test is described in Section \ref{delays}. It requires a global network of sensors with individual stations sensitive to axions with the frequency determined by $m_a$. It  also requires the network to be configured in such a way that it contains two or more nearby stations with a distance of $\sim 100$ km or less between them. We argue that such stations should observe    correlated amplified signals with an event rate of $\sim 0.2/{\rm day}$ and with a time delay (\ref{eq:Delta t}) on the order of a second or less (depending on the actual distance separation between   stations). The presence of such correlation may be a  decisive tool in discriminating the signal from the noise background. 
 
 The estimates are based on the AQN model. 
 Why should one take  this  model seriously? 
 A simple answer is as follows.   Originally, this model  was invented to  explain 
 the observed relation $\Omega_{\rm DM}\sim \Omega_{\rm visible}$  where  the   ``baryogenesis" framework   is replaced with a ``charge-separation" paradigm,  as reviewed  in the Introduction. 
 This model is shown to be   consistent with all available cosmological, astrophysical, satellite and ground-based constraints, where AQNs could leave a detectable electromagnetic signature as reviewed in the Introduction, with one and the  {\it same set }  of parameters. 
 \exclude{Furthermore, it  has been also shown   that the AQNs   could be formed and  could  survive the   early Universe's unfriendly environment. Therefore, the AQNs are  entitled to  serve as the DM candidates by all standards.   Finally, the same AQN framework    may also explain a number of other (naively unrelated) observed phenomena such as excess of the galactic diffuse emission in different frequency bands, the so-called ``Primordial Lithium Puzzle",  ``The Solar Corona Mystery", and   the DAMA/LIBRA puzzling annual modulation,  see Introduction  for the references.}  
 The AQN-induced flux (\ref{flux}) is unambiguously  predicted using the {\it same set } of physical parameters.  The use of the modulations  (\ref{eq:annual}) and  (\ref{eq:daily}) and time delays  (\ref{eq:Delta t})  discussed in this work may reveal  the traces of the AQN {\it directly}, in  contrast with {\it indirect} observations mentioned in the Introduction. 
 
  Finally, we note that in this work we considered detecting the AQNs via the axions that they emit interacting with the Earth. Considering that AQNs also produce considerable amount of energy  from the annihilation with the Earth's baryons, it will likely  be easier to detect the AQN via the associated energy-deposition, for instance, acoustic, signatures\footnote{Electromagnetic signatures in the form of x-rays or $\gamma$-rays will be absorbed  
  in the atmosphere and underground, which makes it hard to detect them. It would be also difficult to discriminate the AQN-induced photons from the background of radiation on the Earth's surface.}. An important point is that the network and modulation approaches discussed above will be helpful in this case as well. We leave this topic for future studies.

\section*{Acknowledgments}
  The work of DB supported in part by the DFG Project ID 390831469:  EXC 2118 (PRISMA+ Cluster of Excellence). DB also received support from the European Research Council (ERC) under the European Union Horizon 2020 Research and Innovation Program (grant agreement No. 695405), from the DFG Reinhart Koselleck Project and the Heising-Simons Foundation.
 The work of  VF is supported by the Australian Research Council, the Gutenberg Fellowship and the New Zealand Institute for Advanced Study.
  This work of AZ and XL was supported in part by the National Science and Engineering Research Council of Canada.

\FloatBarrier
 
\bibliography{AQN_GN}

\end{document}